\documentclass[longauth,useAMS,usenatbib]{mn2e}
\usepackage{graphicx,bm,setspace}
\newcommand{\mymat}[1]{\bm{\mathsf{#1}}}
\newcommand{\myvec}[1]{\bm{#1}}

\title[Lunar occultation of the diffuse radio sky]
  {Lunar occultation of the diffuse radio sky: LOFAR measurements between 35 and 80~MHz} 
\author[Vedantham et al.]
{\parbox{\textwidth}{\fontsize{12pt}{0.5em}\selectfont
{
H.~K.~Vedantham$^{1}$\thanks{E-mail: harish@astro.rug.nl},
L.~V.~E.~Koopmans$^{1}$, 
A.~G.~de Bruyn$^{1,2}$, 
S.~J.~Wijnholds$^{2}$, 
M.~Brentjens$^{2}$, 
F.~B.~Abdalla$^{3}$, 
K.~M.~B.~Asad$^{1}$, 
G.~Bernardi$^{4,5,6}$, 
S.~Bus$^{1}$, 
E.~Chapman$^{3}$, 
B.~Ciardi$^{7}$, 
S.~Daiboo$^{1}$, 
E.~R.~Fernandez$^{1}$, 
A.~Ghosh$^{1}$, 
G.~Harker$^{3}$, 
V.~Jelic$^{1}$, 
H.~Jensen$^{8}$, 
S.~Kazemi$^{1}$, 
P.~Lambropoulos$^{1}$, 
O.~Martinez-Rubi$^{1}$, 
G.~Mellema$^{8}$, 
M.~Mevius$^{1}$, 
A.~R.~Offringa$^{9}$, 
V.~N.~Pandey$^{1}$, 
A.~H.~Patil$^{1}$, 
R.~M.~Thomas$^{1}$, 
V.~Veligatla$^{1}$, 
S.~Yatawatta$^{2}$, 
S.~Zaroubi$^{1}$, 
J.~Anderson$^{10,11}$, 
A.~Asgekar$^{2,12}$, 
M.~E.~Bell$^{13}$, 
M.~J.~Bentum$^{2,14}$, 
P.~Best$^{15}$, 
A.~Bonafede$^{16}$, 
F.~Breitling$^{11}$, 
J.~Broderick$^{17}$, 
M.~Br\"uggen$^{16}$, 
H.~R.~Butcher$^{9}$, 
A.~Corstanje$^{18}$, 
F.~de Gasperin$^{16}$, 
E.~de Geus$^{2,19}$, 
A.~Deller$^{2}$, 
S.~Duscha$^{2}$, 
J.~Eisl\"offel$^{20}$, 
D.~Engels$^{21}$, 
H.~Falcke$^{18,2}$, 
R.~A.~Fallows$^{2}$, 
R.~Fender$^{22}$, 
C.~Ferrari$^{23}$, 
W.~Frieswijk$^{2}$, 
M.~A.~Garrett$^{2,24}$, 
J.~Grie\ss{}meier$^{25,26}$, 
A.~W.~Gunst$^{2}$, 
T.~E.~Hassall$^{17}$, 
G.~Heald$^{2}$, 
M.~Hoeft$^{20}$, 
J.~H\"orandel$^{18}$, 
M.~Iacobelli$^{24}$, 
E.~Juette$^{27}$, 
V.~I.~Kondratiev$^{2,28}$, 
M.~Kuniyoshi$^{29}$, 
G.~Kuper$^{2}$, 
G.~Mann$^{11}$, 
S.~Markoff$^{30}$, 
R. McFadden$^{2}$, 
D.~McKay-Bukowski$^{31,32}$, 
D.~D.~Mulcahy$^{29}$, 
H.~Munk$^{2}$, 
A.~Nelles$^{18}$, 
M.~J.~Norden$^{2}$, 
E.~Orru$^{2}$, 
M.~Pandey-Pommier$^{33}$, 
R.~Pizzo$^{2}$, 
A.~G.~Polatidis$^{2}$, 
W.~Reich$^{29}$, 
A.~Renting$^{2}$, 
H.~R\"ottgering$^{24}$, 
D.~Schwarz$^{34}$, 
A.~Shulevski$^{1}$, 
O.~Smirnov$^{5,4}$, 
B.~W.~Stappers$^{35}$, 
M.~Steinmetz$^{11}$, 
J.~Swinbank$^{30}$, 
M.~Tagger$^{25}$, 
Y.~Tang$^{2}$, 
C.~Tasse$^{36}$, 
S.~ter Veen$^{18}$, 
S.~Thoudam$^{18}$, 
C.~Toribio$^{2}$, 
C.~Vocks$^{11}$, 
M.~W.~Wise$^{2,30}$, 
O.~Wucknitz$^{29}$ and 
P.~Zarka$^{36}$ \\
{\emph{\small{\singlespacing
$^{1}$Kapteyn Astronomical Institute, PO Box 800, 9700 AV Groningen, The Netherlands 
\\
$^{2}$Netherlands Institute for Radio Astronomy (ASTRON), Postbus 2, 7990 AA Dwingeloo, The Netherlands 
\\
$^{3}$Department of Physics and Astronomy, University College London, Gower Street, London WC1E 6BT, UK 
\\
$^{4}$SKA South Africa, 3rd Floor, The Park, Park Road, Pinelands, 7405, South Africa 
\\
$^{5}$Department of Physics and Elelctronics, Rhodes University, PO Box 94, Grahamstown 6140, South Africa 
\\
$^{6}$Harvard-Smithsonian Center for Astrophysics, 60 Garden Street, Cambridge, MA 02138, USA 
\\
$^{7}$Max Planck Institute for Astrophysics, Karl Schwarzschild Str. 1, 85741 Garching, Germany 
\\
$^{8}$Department of Astronomy and Oskar Klein Centre, Stockholm University, AlbaNova, SE-10691 Stockholm, Sweden 
\\
$^{9}$Research School of Astronomy and Astrophysics, Australian National University, Mt Stromlo Obs., via Cotter Road, Weston, A.C.T. 2611, Australia 
\\
$^{10}$Helmholtz-Zentrum Potsdam, DeutschesGeoForschungsZentrum GFZ, Department 1: Geodesy and Remote Sensing, Telegrafenberg, A17, 14473 Potsdam, Germany 
\\
$^{11}$Leibniz-Institut f\"{u}r Astrophysik Potsdam (AIP), An der Sternwarte 16, 14482 Potsdam, Germany 
\\
$^{12}$Shell Technology Center, Bangalore, India 
\\
$^{13}$CSIRO Australia Telescope National Facility, PO Box 76, Epping NSW 1710, Australia 
\\
$^{14}$University of Twente, The Netherlands 
\\
$^{15}$Institute for Astronomy, University of Edinburgh, Royal Observatory of Edinburgh, Blackford Hill, Edinburgh EH9 3HJ, UK 
\\
$^{16}$University of Hamburg, Gojenbergsweg 112, 21029 Hamburg, Germany 
\\
$^{17}$School of Physics and Astronomy, University of Southampton, Southampton, SO17 1BJ, UK 
\\
$^{18}$Department of Astrophysics/IMAPP, Radboud University Nijmegen, P.O. Box 9010, 6500 GL Nijmegen, The Netherlands 
\\
$^{19}$SmarterVision BV, Oostersingel 5, 9401 JX Assen 
\\
$^{20}$Th\"{u}ringer Landessternwarte, Sternwarte 5, D-07778 Tautenburg, Germany 
\\
$^{21}$Hamburger Sternwarte, Gojenbergsweg 112, D-21029 Hamburg 
\\
$^{22}$Astrophysics, University of Oxford, Denys Wilkinson Building, Keble Road, Oxford OX1 3RH 
\\
$^{23}$Laboratoire Lagrange, UMR7293, Universit\`{e} de Nice Sophia-Antipolis, CNRS, Observatoire de la C\'{o}te d'Azur, 06300 Nice, France 
\\
$^{24}$Leiden Observatory, Leiden University, PO Box 9513, 2300 RA Leiden, The Netherlands 
\\
$^{25}$LPC2E - Universite d'Orleans/CNRS 
\\
$^{26}$Station de Radioastronomie de Nancay, Observatoire de Paris - CNRS/INSU, USR 704 - Univ. Orleans, OSUC , route de Souesmes, 18330 Nancay, France 
\\
$^{27}$Astronomisches Institut der Ruhr-Universit\"{a}t Bochum, Universitaetsstrasse 150, 44780 Bochum, Germany 
\\
$^{28}$Astro Space Center of the Lebedev Physical Institute, Profsoyuznaya str. 84/32, Moscow 117997, Russia 
\\
$^{29}$Max-Planck-Institut f\"{u}r Radioastronomie, Auf dem H\"ugel 69, 53121 Bonn, Germany 
\\
$^{30}$Anton Pannekoek Institute, University of Amsterdam, Postbus 94249, 1090 GE Amsterdam, The Netherlands 
\\
$^{31}$Sodankyl\"{a} Geophysical Observatory, University of Oulu, T\"{a}htel\"{a}ntie 62, 99600 Sodankyl\"{a}, Finland 
\\
$^{32}$STFC Rutherford Appleton Laboratory,  Harwell Science and Innovation Campus,  Didcot  OX11 0QX, UK 
\\
$^{33}$Centre de Recherche Astrophysique de Lyon, Observatoire de Lyon, 9 av Charles Andr\'{e}, 69561 Saint Genis Laval Cedex, France 
\\
$^{34}$Fakult\"{a}t fr Physik, Universit\"{a}t Bielefeld, Postfach 100131, D-33501, Bielefeld, Germany 
\\
$^{35}$Jodrell Bank Center for Astrophysics, School of Physics and Astronomy, The University of Manchester, Manchester M13 9PL,UK 
\\
$^{36}$LESIA, UMR CNRS 8109, Observatoire de Paris, 92195   Meudon, France 
}}}
}
}}

\begin{document}
%
\date{\today}
\pagerange{\pageref{firstpage}--\pageref{lastpage}} \pubyear{2012}
\def\LaTeX{L\kern-.36em\raise.3ex\hbox{a}\kern-.15em
    T\kern-.1667em\lower.7ex\hbox{E}\kern-.125emX}
\newtheorem{theorem}{Theorem}[section]
\label{firstpage}
\maketitle
%
%

\begin{abstract}
We present radio observations of the Moon between $35$ and $80$~MHz to demonstrate a novel technique of interferometrically measuring large-scale diffuse emission extending far beyond the primary beam (global signal) for the first time. In particular, we show that (i) the Moon appears as a negative-flux source at frequencies $35<\nu<80$~MHz since it is `colder' than the diffuse Galactic background it occults, (ii) using the (negative) flux of the lunar disc, we can reconstruct the spectrum of the diffuse Galactic emission with the lunar thermal emission as a reference, and (iii) that reflected RFI (radio-frequency interference) is concentrated at the center of the lunar disc due to specular nature of reflection, and can be independently measured. Our RFI measurements show that (i) Moon-based Cosmic Dawn experiments must design for an Earth-isolation of better than $80$~dB to achieve an RFI temperature $<1$~mK, (ii) Moon-reflected RFI contributes to a dipole temperature less than $20$~mK for Earth-based Cosmic Dawn experiments, (iii) man-made satellite-reflected RFI temperature exceeds $20$~mK if the aggregate cross section of visible satellites exceeds $80$~m$^2$ at $800$~km height, or $5$~m$^2$ at $400$~km height. Currently, our diffuse background spectrum is limited by sidelobe confusion on short baselines (10-15\% level). Further refinement of our technique may yield constraints on the redshifted global $21$-cm signal from Cosmic Dawn ($40>z>12$) and the Epoch of Reionization ($12>z>5$).
\end{abstract}
%

\section{Introduction}
The cosmic radio background at low frequencies ($<200$~MHz) consists of Galactic and Extragalactic synchrotron and free-free emission (hundreds to thousands of Kelvin brightness) and a faint (millikelvin level) redshifted $21$-cm signal from the dark ages ($1100>z>40$), Cosmic Dawn ($40>z>15$), and the Epoch of Reionization ($15>z>5$). The redshifted $21$-cm signal from Cosmic Dawn and the Epoch of Reionization is expected to place unprecedented constraints on the nature of the first stars and black-holes \citep{furlanetto2006, pritchard2010}. To this end, many low-frequency experiments are proposed, or are underway: (i) interferometric experiments that attempt to measure the spatial fluctuations of the redshifted $21$-cm signal, and (ii) single-dipole (or total power) experiments that attempt to measure the sky averaged (or global) brightness of the redshifted $21$-cm signal.\\

An accurate single-dipole (all-sky) spectrometer can measure the cosmic evolution of the global $21$-cm signal, and place constraints on the onset and strength of Ly$\alpha$ flux from the first stars and X-ray flux from the first accreting compact objects \citep{mirocha2013}. However, Galactic foregrounds in all-sky spectra are expected to be $4-5$ orders of magnitude brighter than the cosmic signal. Despite concerted efforts \citep{aaron2009,rogers2012,harker2012,patra2013,scihi}, accurate calibration to achieve this dynamic range remains elusive. Single-dipole spectra are also contaminated by receiver noise which has non-thermal components with complex frequency structure that can confuse the faint cosmic signal. The current best upper limits on the 21-cm signal from Cosmic Dawn comes from the SCI-HI group, who reported systematics in data at about 1-2 orders of magnitude above the expected cosmological signal \citep{scihi}. In this paper, we take the first steps towards developing an alternate technique of measuring an all-sky (or global) signal that circumvents some of the above challenges.\\

Multi-element telescopes such as LOFAR \citep{lofar} provide a large number of constraints to accurately calibrate individual elements using bright compact astrophysical sources that have smooth synchrotron spectra. Moreover, interferometric data contains correlations between electric fields measured by independent receiver elements, and is devoid of receiver noise bias. These two properties make interferometry a powerful and well established technique at radio frequencies \citep{thomps}. However, interferometers can only measure fluctuations in the sky brightness (on different scales) and are insensitive to a global signal. However, lunar occultation introduces spatial structure on an otherwise featureless global signal. Interferometers are sensitive to this structure, and can measure a global signal with the lunar brightness as a reference \citep{shaver1999}. At radio frequencies, the Moon is expected to be a spectrally-featureless black body \citep{krotikov1963,heiles1963} making it an effective temperature reference. Thus, lunar occultation allows us to use the desirable properties of interferometry to accurately measure a global signal that is otherwise inaccessible to traditional interferometric observations.\\

The first milestone for such an experiment would be a proof-of-concept, wherein one detects the diffuse or global component of the Galactic synchrotron emission by interferometrically observing its occultation by the Moon. \citet{mckinley2013} recently lead the first effort in this direction at higher frequencies ($80<\nu<300$~MHz) corresponding to the Epoch of Reionization. However, \citet{mckinley2013} concluded that the apparent temperature of the Moon is contaminated by reflected man-made interference (or Earthshine) --- a component they could not isolate from the Moon's intrinsic thermal emission, and hence, they could not estimate the Galactic synchrotron background spectrum. \\

In this paper, we present the first radio observations of the Moon at frequencies corresponding to the Cosmic Dawn ($35<\nu<80$~MHz; $40>z>17$). We show that (i) the Moon appears as a source with negative flux density ($\sim-25$~Jy at $60$~MHz) at low frequencies as predicted, (ii) that reflected Earthshine is mostly specular in nature, due to which, it manifests as a compact source at the center of the lunar disc, and (iii) that reflected Earthshine can be isolated as a point source, and removed from lunar images using the resolution afforded by LOFAR's long baselines ($> 100\lambda$). Consequently, by observing the lunar occultation  of the diffuse radio sky, we demonstrate simultaneous measurement of the spectrum of (a) diffuse Galactic emission and (b) the Earthshine reflected by the Moon. These results demonstrate a new and exiting observational channel for measuring large-scale diffuse emission interferometrically.\\

The rest of the paper is organized as follows. In Section \ref{sec:masbai}, we describe the theory behind extraction of the brightness temperature of diffuse emission using lunar occultation. In Section \ref{sec:proof_of_concept}, we present the pilot observations, data reduction, and lunar imaging pipelines. In Section \ref{sec:analysis}, we demonstrate modeling of lunar images for simultaneous estimation of the spectrum of (i) reflected Earthshine flux, and (ii) the occulted Galactic emission (global signal). We then use the Earthshine estimates to compute limitations to Earth and Moon-based single-dipole Cosmic Dawn experiments. Finally, in Section \ref{sec:concl}, we present the salient conclusions of this paper, and draw recommendations for future work.
%
%
%
%
\section{Lunar occultation as seen by a radio interferometer}
\label{sec:masbai}
As pointed out by \citet{shaver1999}, the Moon may be used in two ways to estimate the global $21$-cm signal. Firstly, since the Moon behaves as a $\sim 230$~K thermal black-body at radio frequencies \citep{krotikov1963,heiles1963}, it can be used in place of a man-made temperature reference for bandpass calibration of single-dish telescopes. Secondly, the Moon can aid in an interferometric measurement of the global signal, which in the absence of the Moon would be resolved out by the interferometer baselines. In this section, we describe the theory and intuition behind this quirk of radio interferometry. For simplicity, we will assume that the occulted sky is uniformly bright (global signal only), and that the primary beam is large compared to the Moon, so that we can disregard its spatial response.
%
%
%
%
\subsection{Interferometric response to a global signal}
\label{subsec:resp_to_global_signal}
A radio interferometer measures the spatial coherence of the incident electric field. We denote this spatial coherence, also called visibility, at frequency $\nu$, on a baseline separation vector $\bar{u}$ (in wavelength units) as $V(\bar{u},\nu)$. The spatial coherence function $V(\bar{u},\nu)$ is related to the sky brightness temperature $T(\bar{r},\nu)$ through a Fourier transform given by \citep{thomps}
\begin{equation}
\label{eqn:me}
V(\bar{u},\nu) = \frac{1}{4\pi}\int\,\,d\Omega\,T_{{sky}}(\bar{r},\nu)\,e^{-2\pi\mathrm{i}\bar{u}.\bar{r}},
\end{equation}
where $\bar{r}$ is the unit vector denoting direction on the sky, and $d\Omega$ is the differential solid angle, and we have assumed an isotropic antenna response. Note that we have expressed the visibility in terms of brightness temperature rather than flux density, since brightness temperature gives a more intuitive understanding of a diffuse signal such as the global $21$-cm signal as compared to the flux density which lends itself to representation of emission from compact sources.\\

If the sky is uniformly bright as in the case of a global signal, then\footnote{In this context, we use $T_B(\nu)$ to denote a generic spatially invariant function--- not necessarily the particular case of the global $21$-cm signal.} $T_{{sky}}(\bar{r},\nu)=T_B(\nu)$, and by using a spherical harmonic expansion of the integrand in Equation \ref{eqn:me}, we can show that (full derivation in Appendix \ref{app:a})
\begin{equation}
\label{eqn:global}
V(\bar{u}, \nu) = T_B(\nu)\,\frac{\sin(2\pi|\bar{u}|)}{2\pi|\bar{u}|}
\end{equation}
where $|\bar{u}|$ is the length of vector $\bar{u}$. Equation \ref{eqn:global} implies that the response of an interferometer with baseline vector $\bar{u}$ (in wavelengths) to a global signal (no spatial structure) falls off as $\sim 1/|\bar{u}|$. Hence only a zero baseline, or very short baselines ($<$~few $\lambda$) are sensitive to a global signal. This is seen in Figure \ref{fig:two_resp} (solid red curve) where we plot the interferometric response to a global signal as given by Equation \ref{eqn:global}.\\

In zero-baseline (single-dipole) measurements, the signal is contaminated by receiver noise which is very difficult to measure or model with accuracy ($\sim 10$~mK level). This is evident from recent efforts by \citet{aaron2009,rogers2012,harker2012,patra2013,scihi}. On shorter non-zero baselines, the receiver noise from the two antennas comprising the baseline, being independent, has in principle, no contribution to the visibility $V(\bar{u})$. In practice however, short baselines ($<$ few wavelengths) are prone to contamination from mutual coupling between the antenna elements, which is again very difficult to model. Moreover, non-zero baselines are also sensitive to poorly constrained large-scale Galactic foregrounds and may not yield accurate measurements of a global signal.
%
%
%
%
\subsection{Response to occultation}
\label{subsec:resp_to_occultation}
The presence of the Moon in the field of view modifies $T_{{sky}}(\nu)$. If the lunar brightness temperature is $T_M(\nu)$ and the global brightness temperature in the field is $T_B(\nu)$, then lunar occultation imposes a disc-like structure of magnitude $T_M(\nu)-T_B(\nu)$ on an otherwise featureless background.\\

To mathematically represent the occultation, we introduce the masking function which is unity on a disc the size of the Moon ($\sim 0.5$~deg) centered on the Moon at $\bar{r}_M$, and zero everywhere:
%
%
%
\begin{equation}
\label{eqn:mask}
M(\bar{r},\bar{r_M}) = \left\{ \begin{array}{rl} 
	1 & \bar{r}.\bar{r}_M > \cos(0.25\textrm{~deg})\\
	0 & \textrm{otherwise.} 
	\end{array} \right\}
\end{equation}
The sky brightness may then be expressed as,
\begin{equation}
\label{eqn:sky_with_mask}
\begin{array}{rcl}
T_{{sky}}	&=& T_B\,(1-M)+T_M\,M \\
			&=& \underbrace{(T_M-T_B)M}_{\textrm{occulted}}\,\,\,\, + \underbrace{T_B}_{\textrm{non-occulted}},
\end{array}
\end{equation}
where the function arguments have been dropped for brevity. Equation \ref{eqn:sky_with_mask} shows that $T_{{sky}}$ consists of (i) a spatially fluctuating `occulted' component\footnote{Though $T_B$ and $T_M$ are assumed here to have no spatial fluctuations, $(T_M-T_B)M$ has spatial fluctuations due to $M$.}: $(T_M-T_B)M$, and (ii) a spatially invariant `non-occulted' component: $T_B$ akin to the occultation-less case discussed in Section \ref{subsec:resp_to_global_signal}. By substituting Equation \ref{eqn:sky_with_mask} in Equation \ref{eqn:me}, we get an expression for the visibility:
\begin{eqnarray}
\label{eqn:measured_vis}
V(\bar{u}) &=& \underbrace{\frac{1}{4\pi}(T_M-T_B)\int\,d\Omega\,M(\bar{r},\bar{r}_M)\,e^{-2\pi\mathrm{i}\bar{u}.\bar{r}}}_{\textrm{occulted}} \nonumber \\
           &+& \underbrace{\frac{1}{4\pi}T_B\int\,d\Omega\,e^{-2\pi\mathrm{i}\bar{u}.\bar{r}}.}_{\textrm{non-occulted}}
\end{eqnarray}
The above equation is central to the understanding of the occultation based technique presented in this paper. The second integral representing the `non-occulted' brightness evaluates to a sinc-function (see Equation \ref{eqn:global}):
\begin{equation}
\label{eqn:constant}
V_2(\bar{u}) = T_B\frac{\sin(2\pi|\bar{u}|)}{2\pi|\bar{u}|}\,\,\,\textrm{(non-occulted)}
\end{equation}
This is the interferometric response we expect in the absence of any occultation.\\

The first integral is the additional term generated by the occultation. It is a measure of $T_B-T_M$: the differential brightness temperature between the background being occulted and the occulting object itself. This integral is simply the Fourier transform of a unit disc, and if we (i) assume that the angular size of the occulting object, $a$ is small, and (ii) use a co-ordinate system which has its $z$-axis along $\bar{r}_M$, then the first integral may be approximated to a sinc-function:
\begin{equation}
\label{eqn:fluctuating}
V_1(\bar{u}) \approx (T_M-T_B)\,\frac{\sin(\pi a|\bar{u}|)}{\pi a|\bar{u}|}\,\,\,\textrm{(occulted)}.
\end{equation}
This is the additional interferometric response due to the presence of the occulting object.\\

\begin{figure}
\includegraphics[width=\linewidth]{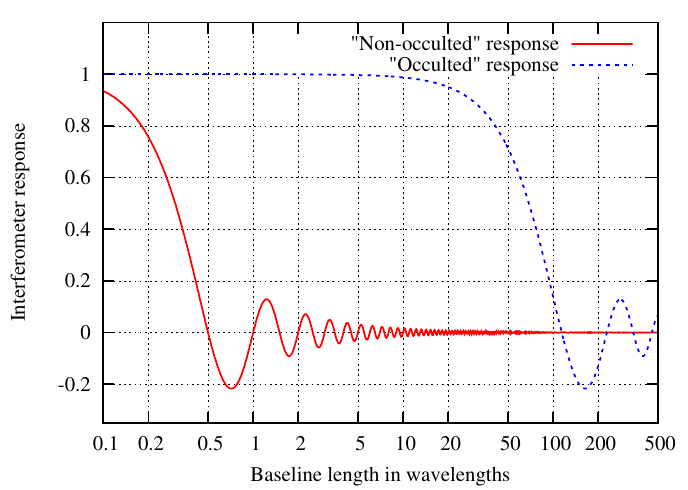}
\caption{Response of an interferometer to lunar occultation as a function of baseline length: the blue broken line shows the `occulted' component (Equation \ref{eqn:fluctuating}) and the solid red line shows the `non-occulted' component (Equation \ref{eqn:constant}). The solid red line is also the interferometer response in the absence of occultation. \label{fig:two_resp}}
\end{figure}

In Figure \ref{fig:two_resp}, we plot the interferometric response to the `occulted' and `non-occulted' components as a function of baseline length for $a=0.5$~deg. Clearly, longer baselines ($|\bar{u}|\sim 1/a$) of a few tens of wavelengths are sensitive to the `occulted' component $T_M-T_B$. With prior knowledge of $T_M$, the global background signal $T_B$ may be recovered from these baselines. The longer baselines also have two crucial advantages: (i) they contain negligible mutual coupling contamination, and (ii) they probe scales on which Galactic foreground contamination is greatly reduced as compared to very short baselines (few wavelengths). These two reasons are the primary motivations for the pilot project presented in this paper.
%
%
%
%
%
%
\subsection{Lunar brightness: a closer look}
\label{subsec:lunar_brightness}
\begin{figure}
\centering
\includegraphics[width=\linewidth]{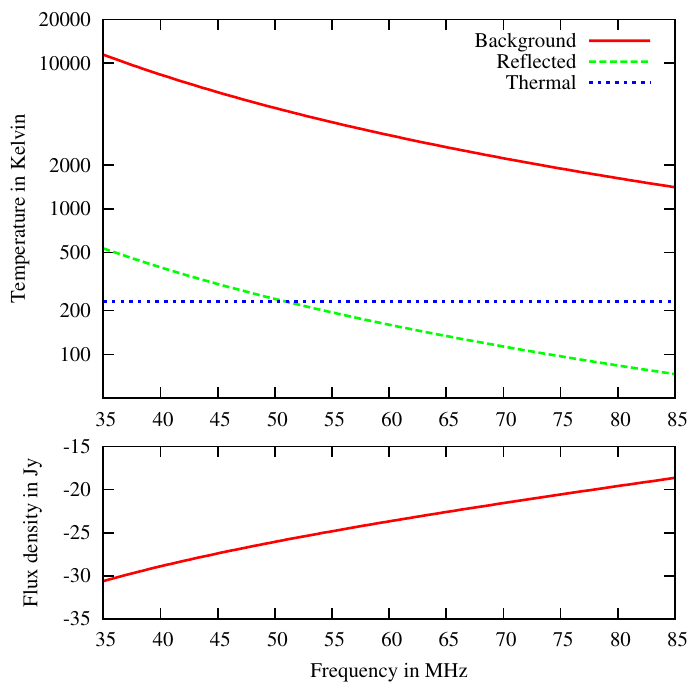}
\caption{Plot showing the expected temperature of occulted Galactic synchrotron emission, reflected Galactic emission, and lunar thermal emission as a function of frequency (top panel), and also the resulting flux density of the Moon as measured by an interferometer (bottom panel).\label{fig:expected}}
\end{figure}

Since an interferometer measures $T_M-T_B$, our recovery of $T_B$ hinges on our knowledge of $T_M$. We have thus far assumed that the lunar brightness temperature, $T_M$, is given by a perfect black-body spectrum without spatial structure. In practice, the apparent lunar temperature consists of several non-thermal contributions. We now briefly discuss these in decreasing order of significance.
\begin{enumerate}
\item \emph{Reflected RFI: }
Man-made interference can reflect off the lunar disc into the telescope contributing to the effective lunar temperature. Radar studies have shown that the lunar surface appears smooth and undulating at wavelengths larger than $\sim 5$~metres \citep{evans1969}. Consequently, at frequencies of interest to Cosmic Dawn studies ($35<\nu<80$~MHz) Earthshine-reflection is expected to be specular in nature and may be isolated by longer baselines ($>100\,\lambda$) to the center of the lunar disc. This `reflected Earthshine' was the limiting factor in recent observations by \citet{mckinley2013}. In Section \ref{sec:analysis}, we demonstrate how the longer baselines of LOFAR ($>100\,\lambda$) can be used to model and remove Earthshine from images of the Moon. \\
\item \emph{Reflected Galactic emission: }
The Moon also reflects Galactic radio emission incident on it. As argued before, the reflection at tens of MHz frequencies is mostly specular. Radar measurements of the dielectric properties of the lunar regolith have shown that the Moon behaves as a dielectric sphere with an albedo of $\sim 7$\% \citep{evans1969}. If the sky were uniformly bright (no spatial structure) this would imply an additional lunar brightness of $0.07\,T_B(\nu)$~K, leading to a simple correction to account for this effect. In reality, Galactic synchrotron has large-scale structure (due to the Galactic disk) and the amount of reflected emission depends on the orientation of the Earth---Moon vector with respect to the Galactic plane and changes with time. For the sensitivity levels we reach with current data (see Section \ref{subsec:sensitivity_analysis}), it suffices to model reflected emission as a time-independent temperature of $160$~K at $60$~MHz with a spectral index of $-2.24$:
\begin{equation}
T_{refl} = 160\,\left(\frac{\nu\,\,\textrm{MHz}}{60}\right)^{-2.24}
\end{equation}
Details of simulations that used the sky model from \citet{dacosta} to arrive at the above equation are presented in Appendix \ref{app:b}.\\

\item \emph{Reflected solar emission: }
The radio-emitting region of the quiet Sun is about $ 0.7$~deg in diameter, has a disc-averaged brightness temperature of around $10^6$~K at our frequencies, and has a power-law-like variation with frequency \citep{erickson1977}. Assuming specular reflection, the reflected solar emission from the Moon is localized to a $7$~arc-sec wide region on the Moon, and as such can be modeled and removed. In any case, assuming an albedo of $7$\%, this adds a contribution of about $1$~K to the disc-integrated temperature of the Moon. While this contribution must be taken into account for $21$-cm experiments, we currently have not reached sensitivities where this is an issue. For this reason, we discount reflected solar emission from the quiet Sun in this paper. During disturbed conditions, transient radio emission from the Sun may increase by $\sim 4$~ orders of magnitude and have complex frequency structure. According to SWPC\footnote{National Oceanic and Atmospheric Administration's Space Weather Prediction Center (http://www.swpc.noaa.gov)}, such events were not recorded during the night in which our observations were made.\\
\item \emph{Polarization: }
The reflection coefficients for the lunar surface depend on the polarization of the incident radiation. For this reason both the thermal radiation from the Moon, and the reflected Galactic radiation is expected to be polarized. \citet{moffat1972} has measured the polarization of lunar thermal emission to have spatial dependence reaching a maximum value of about $18$\% at the limbs. While leakage of polarized intensity into Stokes I brightness due to imperfect calibration of the telescopes is an issue for deep $21$-cm observations, we currently do not detect any polarized emission from the lunar limb in our data. The most probable reason for this is depolarization due to varying ionospheric rotation measure during the synthesis, and contamination from imperfect subtraction of a close ($\sim 5$~deg), polarized, and extremely bright source (Crab pulsar).
\end{enumerate}
In this paper, we model the lunar brightness temperature as a sum of its intrinsic black-body emission, and reflected emission:
\begin{equation}
\label{eqn:tmoon}
T_M = T_{black}+T_{refl} = 230 + 160\,\left(\frac{\nu\,\,\textrm{MHz}}{60}\right)^{-2.24}\,\, K
\end{equation}
The top-panel in Fig. \ref{fig:expected} shows the expected diffuse background $T_B$, and the above two contributions to the lunar brightness temperature $T_M$ as a function of frequency assuming that the Moon is located at $05$h$13$m$00$s, $+20$d$26$m$00$s\footnote{This corresponds to the transit point of the Moon during the observations presented in this paper}. The bottom panel of Fig. \ref{fig:expected} also shows the flux density of the lunar disc $S_m$ (Stokes I) given by
\begin{equation}
\label{eqn:stot}
S_{m} = \frac{2k(T_M-T_B)\Omega}{\lambda^2 10^{-26}} \,\,\,\textrm{Jy},
\end{equation}
where the temperatures are expressed in Kelvin, $k$ is the Boltzmann's constant, $\lambda$ is the wavelength, and $\Omega$ is the solid angle subtended by the Moon. Since $T_M-T_B$ is negative in our frequency range, the Moon is expected to appear as a negative source with flux density of about $-25$~Jy at $60$~MHz. In Section \ref{subsec:final_estimation}, we will use Equations \ref{eqn:stot} and \ref{eqn:tmoon}, along with a measurement of the $S_{m}$, to compute the spectrum of the diffuse background $T_B$.
%
%
%
%
\section{Proof of concept}
\label{sec:proof_of_concept}
%
%
%
We acquired $7$ hours of LOFAR \citep{lofar} commissioning data between 26-12-2012 19:30 UTC and 27-12-2012 02:30 UTC. $24$ Low Band Antenna (LBA) core-stations (on a common clock) and $11$ remote stations participated in the observation\footnote{Station here refers to a phased array of dipoles that forms the primary antenna element in the interferometer}. The core-stations are distributed within a $\sim 3$~km core near the town of Exloo in the Netherlands, and the remote-stations are distributed up to $\sim 50$~km away from the core within the Netherlands. Visibility data were acquired in two simultaneous (phased array) primary beams : (i) a calibration beam on 3C123 (04h37m04s,+29d40m13.8s), and (ii) a beam in the direction of the Moon at transit (05h13m, +20d26m)\footnote{All co-ordinates are specified for equinox J2000 at epoch J2000, but for lunar co-ordinates that are specified for equinox J2000 at epoch J2012 Dec 26 23:00 UTC}. We acquired data on the same $244$~sub-bands (each $\sim 195$~kHz wide) in both beams. These subbands together span a frequency range from $\sim 36$~MHz to $\sim 83.5$~MHz. The raw data were acquired at a time/freq resolution of $1$~sec, $3$~kHz ($64$~channels per sub-band) to reduce data loss to radio-frequency interference (RFI), giving a raw-data volume of about $ 15$~Terabytes.
%
%
%
%
%
\subsection{RFI flagging and calibration}
\label{subsec:rfi_flagging}
RFI ridden data-points were flagged using the \texttt{AOFlagger} algorithm \citep{aoflagger} in both primary (phased-array) beams. The 3C123 calibrator beam data were then averaged to $5$~sec, $195$~kHz ($1$~channel per sub-band) resolution such that the clock-drift errors on the remote stations could still be solved for. We then bandpass calibrated the data using a two component model for 3C123. The bandpass solutions for the core-station now contain both instrumental gains and ionospheric phases in the direction of 3C123. Since 3C123 is about $ 12$~deg from the Moon, the ionospheric phases may not be directly translated to the lunar field. To separate instrumental gain and short-term ionospheric phase, we filtered the time-series of the complex gains solutions in the Fourier domain (low-pass) and then applied them to the lunar field. 
%
%
%
%
%
\subsection{Subtracting bright sources}
\label{subsec:sub_bright_sources}
%
%
%
%
%
%
%
\begin{figure*}
\centering
\includegraphics[width=\linewidth]{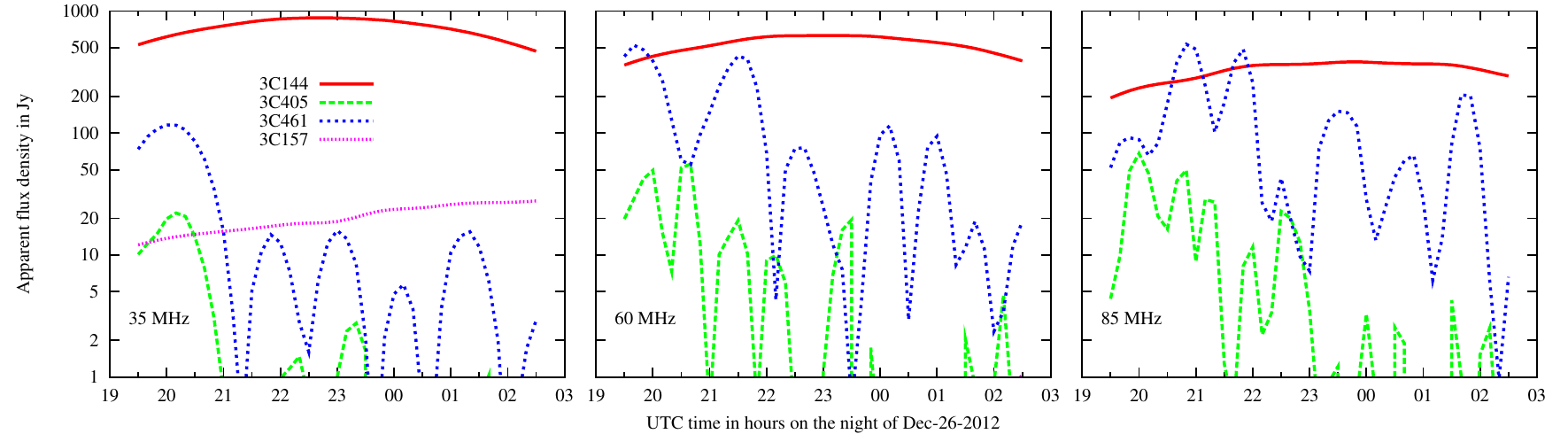}
\caption{Plot showing the simulated apparent flux (primary-beam attenuated) of sources which dominate the flux budget in the lunar field for three different frequencies: $35$~MHz, $60$~MHz, and $80$~MHz. The modulations in time are due to the sources moving through the sidelobes of the primary (station) beam during the synthesis.\label{fig:app_flux}}
\end{figure*}
The brightest source in the lunar field is 3C144 which is about $5$~deg from the pointing center (see Fig. \ref{fig:moon_field}). 3C144 consists of a pulsar ($250$~Jy at $60$~MHz) and a nebula ($2000$~Jy at $60$~MHz) which is around $4$~arcmin wide. 
After RFI flagging, the lunar field data were averaged to $2$~sec, $13$~kHz ($15$~channel per sub-band). This ensures that bright sources away from the field such as Cassiopeia A  are not decorrelated, such that we may solve in their respective directions for the primary beam and ionospheric phase, and remove their contribution from the data. We used the third Cambridge Catalog \citep{3ccat} with a spectral index scaling of $\alpha=-0.7$, in conjunction with a nominal LBA station beam model to determine the apparent flux of bright sources in the lunar beam at $35$, $60$ and $85$~MHz (see Figure \ref{fig:app_flux}). Clearly, 3C144 (Crab nebula) and 3C461 (Cassiopeia A) are the dominant sources of flux in the lunar beam across the observation bandwidth. Consequently, we used the BlackBoard Self-cal (BBS\footnote{BBS is a self-calibration software package developed for LOFAR}) software package to calibrate in the direction of Cassiopeia A, and the Crab nebula simultaneously, and also subtract them from the data. We used a Gaussian nebula plus point source pulsar model for the Crab nebula, and a two component Gaussian model for Cassiopeia A in the solution and subtraction process.  
\subsection{Faint source removal}
\label{subsec:faint_source_removal}
\begin{figure*}
\centering
\includegraphics[width=\linewidth]{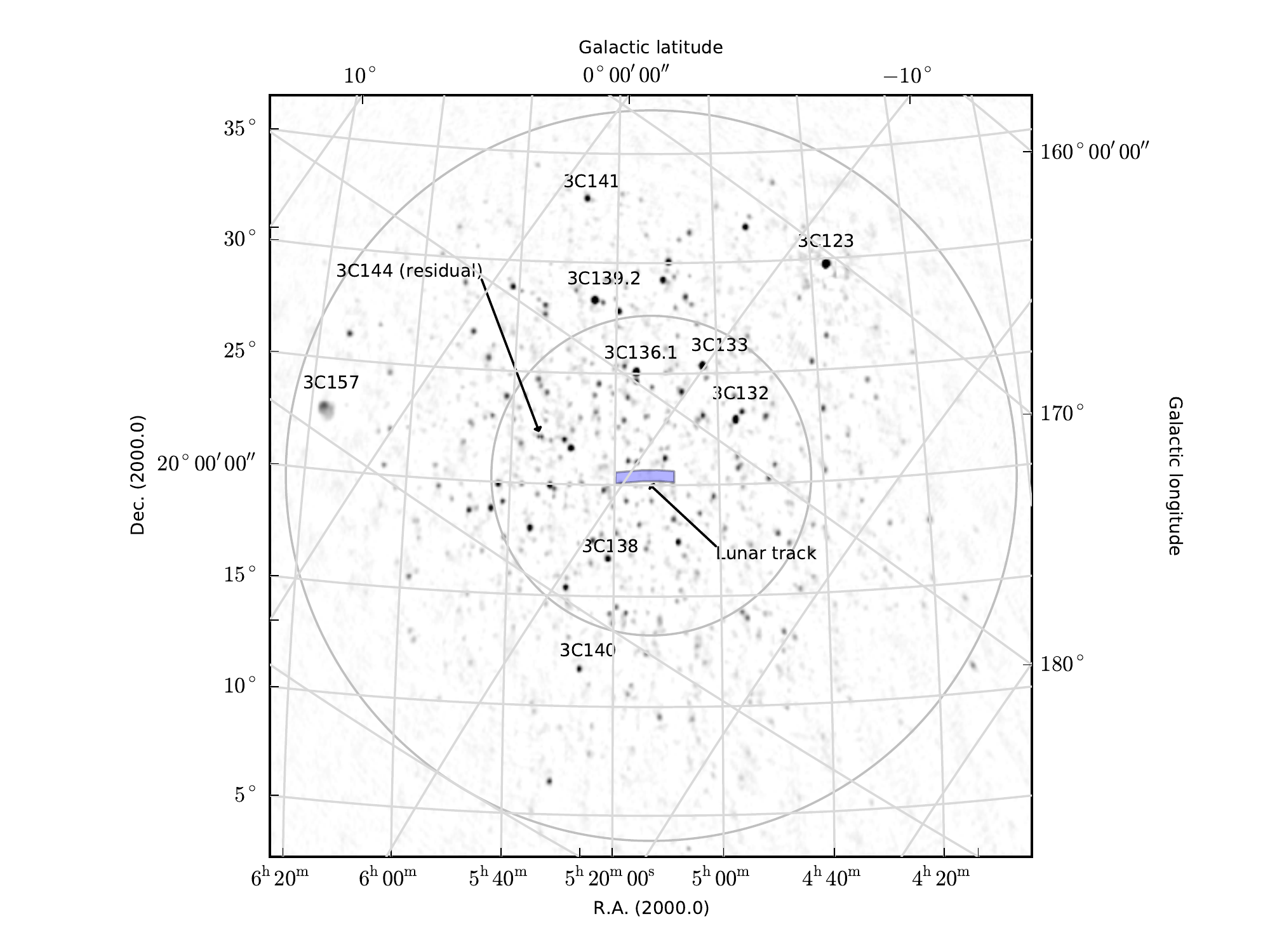}
\caption{Continuum image of the lunar field between $36$ and $44$~MHz (natural weights) with no primary-beam correction applied. The point spread function is about $12$~arcmin wide. Prominent 3C source names have been placed right above the respective sources. 3C144 (Crab nebula) has been subtracted in this image. The gray circles approximately trace the first null of the primary beam at $35$ (outer) and $80$~MHz (inner). The shaded rectangular patch shows the trajectory of the Moon during the $7$~hr synthesis. We have not fringe-stopped on the Moon (moving source) in producing this image, and hence the Moon is not visible due to the ensuing decorrelation.\label{fig:moon_field}}
\end{figure*}
Since the Moon is a moving target (in celestial co-ordinates), all the background sources and their sidelobes will be spatially smeared in lunar images. Standard imaging routines cannot clean (deconvolve) the sidelobes of such spatially smeared sources. To mitigate sidelobe confusion, it is thus important to model and subtract as many background sources as possible prior to lunar imaging. After removing the bright sources, Cassiopeia A and the Crab nebula, we average the data to a resolution of $10$~sec, $40$~kHz ($5$~channel per sub-band) to ensure that the faint sources that populate the relatively large beam (FWHM of $\sim 20$~deg at $35$~MHz) are not decorrelated from time and bandwidth smearing, such that they may be reliably subtracted. We then imaged the sources and extracted a source catalog consisting of $\sim 200$~sources in the field (see Fig. \ref{fig:moon_field}). These sources were clustered into $10-12$~directions depending on frequency. The \texttt{SAGECal} software \citep{SAGECal} was then used to solve for the primary-beam variation in these directions with a solution cadence of $40$~minutes, and subsequently these sources were subtracted to form the visibilities. We note here that none of the sources detected for the $200$~source catalog were occulted by the Moon. In future experiments, if bright sources are occulted by the Moon during the synthesis, then care must be exercised to (i) not subtract them during the occultation, and (ii) flag data at the beginning and end of occultation that contains edge diffraction effects. \\

Multi-direction algorithms may subtract flux from sources that are not included in the source model in their quest to minimize the difference between data and model \citep{robustcal,grobler2014}. Furthermore, the Moon is a moving source (in the celestial co-ordinates frame in which astrophysical sources are fixed). The Moon fringes on a $100\lambda$ East-West baseline at a rate of about $360$~deg per hour. Based on fringe-rate alone, such a baseline will confuse the Moon for a source that is a few degrees from the phase center that also fringes at the same rate. This implies that calibration and source subtraction algorithms are expected to be even more prone to subtracting lunar flux due to confusion with other sources in the field. While a discussion on the magnitude of these effects are beyond the scope of this paper, to mitigate these effects, we excluded all baselines less than $100\lambda$ (that are sensitive to the Moon) in the \texttt{SAGECal} solution process to avoid suppressing flux from the Moon, and also to avoid confusion from the unmodeled diffuse Galactic emission.

\subsection{Lunar imaging}
\label{subsec:lunar_imaging}
\begin{figure*}
\centering
\includegraphics{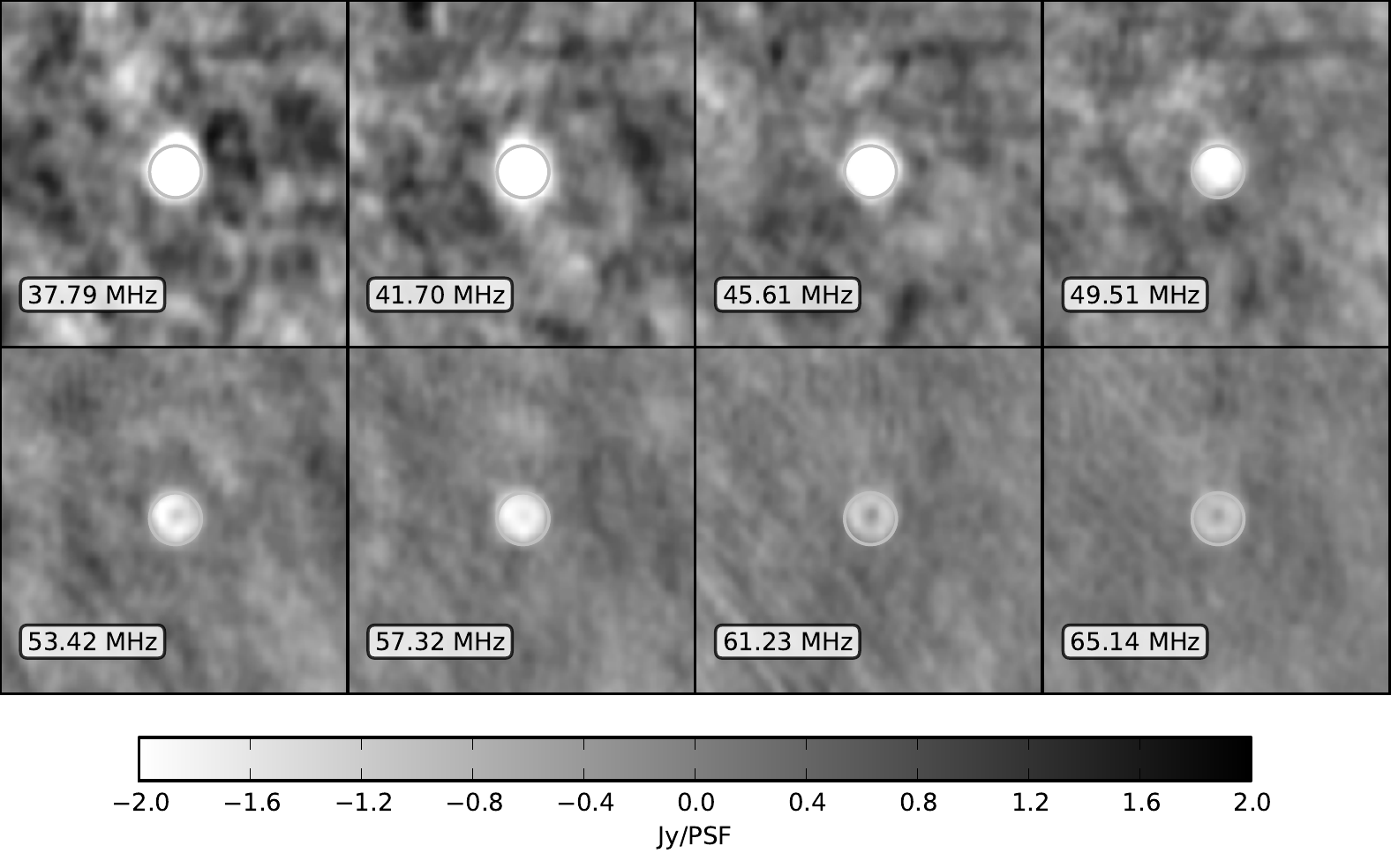}
\caption{Synthesis images of the Moon at $8$ different frequencies: $0.5$~deg wide gray circles are drawn centered on the expected position of the Moon. Each image is made over a bandwidth of $3.9$~MHz ($20$~sub-bands) for $7$~hours of synthesis. The lunar flux increases towards zero with increasing frequency as the contrast between the Galactic background and lunar thermal emission is decreasing (as expected).\label{fig:moon_frames}}
\end{figure*}

After faint source removal, the data were averaged to a resolution of $1$~min, $195$~kHz ($1$~channel per sub-band) to reduce data volume while avoiding temporal decorrelation of the Moon (moving target). We then phase rotated (fringe stopping) to the Moon while taking into account its position and parallax given LOFAR's location at every epoch. After this step, we used standard imaging routines in CASA to make a dirty image of the Moon. Since we are primarily interested in the spectrum of lunar flux in images, it is critical to reduce frequency dependent systematics in the images. As shown later, sidelobe confusion on short baselines ($<100\lambda$) is the current limitation in our data. This contamination is a result of the frequency dependent nature of our native uv-coverage--- an indispensable aspect of Fourier synthesis imaging. To mitigate this effect on shorter baselines, we choose an inner uv-cut of $20\lambda$ for all frequency channels. This value was chosen as it corresponds to the shortest baselines (in wavelengths) available at the highest frequencies in our observation bandwidth. Figure \ref{fig:moon_frames} shows $3.9$~MHz wide ($20$~subbands) continuum images of the Moon in different frequency bands. As expected the Moon appears as a source with negative flux density. As frequency increases, the brightness contrast between the Moon and the background Galactic emission decreases, and the (absolute) flux of the Moon decreases, just as expected.
%
%
%
%
%
%
%
%
\section{Image analysis}
\label{sec:analysis}
In this section, we use the lunar images such as the ones shown in Fig. \ref{fig:moon_frames} to extract the background temperature $T_B$ and the reflected RFI (Earthshine) flux.
\subsection*{Sensitivity analysis}
\label{subsec:sensitivity_analysis}
\begin{figure}
\centering
\includegraphics[width=\linewidth]{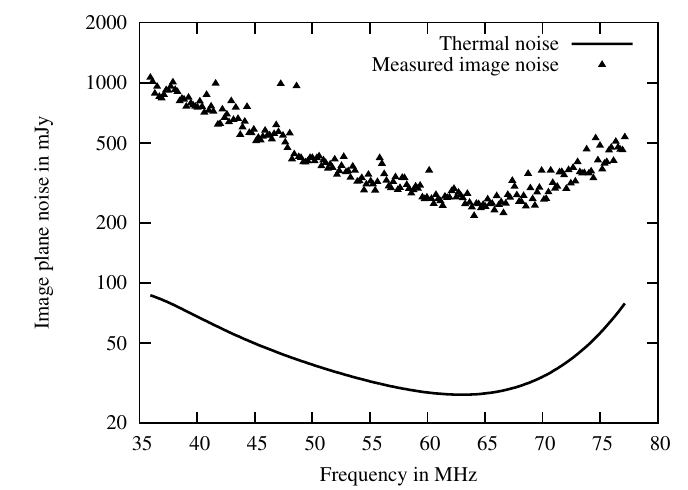}
\caption{Plot showing noise estimated in the lunar images. The thermal noise was estimated by differencing visibilities between frequency channels separated by about $40$~kHz. The measured image noise is shown for naturally weighted images ($20\lambda-500\lambda$~baselines)  made over a bandwidth of about $195$~kHz with a $7$~hour synthesis. \label{fig:image_noise}}
\end{figure}

The noise in lunar images such as the ones in Fig. \ref{fig:moon_frames} primarily originate from (i) thermal noise, (ii) classical and sidelobe confusion noise, and (iii) residuals of the bright in-field source 3C144 (about $2250$~Jy at $60$~MHz)\footnote{The dirty images of unsubtracted sources are spatially smeared in lunar images since we have fringe stopped on the Moon}. To estimate thermal noise, we differenced the bandpass calibrated visibilities between channels separated by $40$~kHz. The aggregated flux from sources across the sky are correlated on such a small frequency interval ($0.01$~per-cent), and drop off in the difference. After taking into account the frequency and time resolution of the visibilities, the rms of the differenced visibilities gives us an independent measurement of the System Equivalent Flux Density (SEFD). From the SEFD, we compute the thermal noise in natural weighted images as
\begin{equation}
\label{eqn:image_noise}
\sigma_{{th}}(\nu) = \frac{\textrm{SEFD}(\nu)}{\sqrt{2\,t_{{syn}}\,\Delta\nu\, N_{{bas}}}},\end{equation}
where $t_{syn}$ is the observation duration, $\Delta \nu$ is the bandwidth, and $N_{bas}$ is the number of baselines used in imaging. Figure \ref{fig:image_noise} shows the expected thermal noise in lunar images (solid black line). The thermal noise is sky limited for $\nu<65$~MHz, and receiver noise begins to become a significant contributor at higher frequencies. Fig. \ref{fig:image_noise} also shows the measured rms noise in Stokes I lunar images (solid triangles), which is a factor of $\sim 10$ higher than the thermal noise alone. The expected confusion noise is about $15$~mJy at $60$~MHz and increases to about $50$~mJy at $35$~MHz, and cannot account for the excess. Simulations of the effects of 3C sources outside the field of view yielded a sidelobe noise of about $100$~mJy at $60$~MHz, which may account for no more than a third of the excess. Most of the remaining excess may be due to the residuals of 3C144 which is only about $5$~degree from the Moon, is scintillating (ionospheric) and is extremely bright ($2200$~Jy at $60$~MHz, and $3600$~Jy at $35$~MHz) and possibly polarized. We currently do not have reliable models to get rid of its contribution with an error less than $0.5$\%. We thus defer a detailed sensitivity analysis of the lunar occultation technique to a future paper where we observe the Moon in a field devoid of extremely bright sources.

\subsection*{Mitigating Earthshine}
\label{subsec:lls_fit}
The lunar images in certain frequency channels comprise of a negative disc (as expected), along with a bright point-source like emission at the center of the disc (see for instance the panels corresponding to $61.23$ and $65.14$~MHz in Fig. \ref{fig:moon_frames}). Since this point-source like emission is always present only at the center of the lunar disc, it is not intrinsic to the Moon. Additionally, it can not emanate from man-made satellites orbiting the Moon, since the apparent positions of man-made satellites must change during the $7$~hour synthesis. The only known source we may attribute this point-source like emission to is Earthshine--- man-made RFI reflected off the lunar surface. Since the lunar surface is expected to be smoothly undulating on spatial scales comparable to the wavelength ($4$ to $9$ meters), reflection off the lunar surface is expected to be mostly specular. Hence Earthshine always images to the center of the lunar disc. The angular size of Earthshine in lunar images may be approximately written as (derivation in Appendix \ref{app:c})
\begin{equation}
\label{eqn:spot}
\Delta \theta_\textrm{es} \approx \frac{R_e\,R_m}{D^2},
\end{equation}
where $R_e$ and $R_m$ are the radius of the Earth and Moon, respectively, and $D$ is the distance between the Earth and the Moon. The approximation holds for $D\gg R_m$ and $D\gg R_e$. Assuming approximate values: $R_e=6400$~km, $R_m=1740$~km, $D=384400$~km, we get $\Delta \theta_\textrm{es} \approx 15.5$~arc-sec: reflected Earthshine is expected to be resolved only on baselines of tens of thousands of wavelengths, and we can safely treat it as a point source for our purposes \footnote{LOFAR's international baselines can achieve sub-arcsecond resolutions. In future, this may be used to make maps of reflected Earthshine!}.\\

We thus model the dirty lunar images with two components: (i) a lunar disc of $0.5$~deg diameter with negative flux $S_m$, and (ii) a point source centered on the lunar disc with positive flux, $S_{es}$. If $\mymat{D}$ is a matrix with dirty image flux values, $\mymat{M}$ is a mask as defined in Equation \ref{eqn:mask}, and $\mymat{P}$ is the telescope Point Spread Function (PSF) matrix then our model may be expressed as 
\begin{equation}
\mymat{D}=\left(S_{m}\mymat{M}+S_{es}\right)\ast \mymat{P} +\mymat{N},
\end{equation}
or equivalently,
\begin{equation}
\mymat{D}=S_{m}\mymat{G}+S_{es}\mymat{P} +\mymat{N}
\end{equation}
where $\mymat{G}=\mymat{M}\ast\mymat{P}$ is the dirty image of the unit disc given the telescope PSF, and $\ast$ denotes $2$-D convolution. The above equation may be vectorized and cast as a linear model with $S_m$ and $S_{es}$ are parameters:
\begin{equation}
\textrm{vec}(\mymat{D}) = \mymat{H}\myvec{\theta}+\textrm{vec}(\mymat{N})
\end{equation} 
where 
\begin{eqnarray}
\mymat{H} &=& [\textrm{vec}(\mymat{G})\,\,\,\textrm{vec}(\mymat{P})] \nonumber \\
\myvec{\theta}&=&[S_{m}\,\,\,S_{es}]^T
\end{eqnarray}
The Maximum Likelihood Estimate of the parameters is then
\begin{equation}
\label{eqn:lls}
\widehat{\myvec{\theta}} = [\widehat{S}_m\,\,\,\widehat{S}_{es}]^T =  (\mymat{H}^T\mymat{H})^{-1}\mymat{H}^T\textrm{vec}(\mymat{D}),
\end{equation}
and the residuals of fitting are given by
\begin{equation}
\textrm{vec}(\mymat{R}) = \textrm{vec}(\mymat{D})-\mymat{H}\widehat{\myvec{\theta}}
\end{equation}
In practice, we solve Equation \ref{eqn:lls} with a positivity constraint on $\widehat{S}_{es}$, and a negativity constraint on $\widehat{S}_m$. Finally, if $\sigma^2_N$ is the noise variance, the parameter covariance matrix is given by
\begin{equation}
\label{eqn:cov}
\mathrm{cov}(\myvec{\theta}) = \sigma^2_N(\mymat{H}^T\mymat{H})^{-1}
\end{equation}
Note that we have assumed that the noise covariance matrix is diagonal with identical entries along its diagonal. A more realistic noise covariance matrix may be estimated from autocorrelation of the images themselves, but we found this to lead to marginal change in the background spectrum which is still dominated by larger systematic errors (see Figure \ref{fig:tback}). \\

\begin{figure*}
\includegraphics[width=0.75\linewidth]{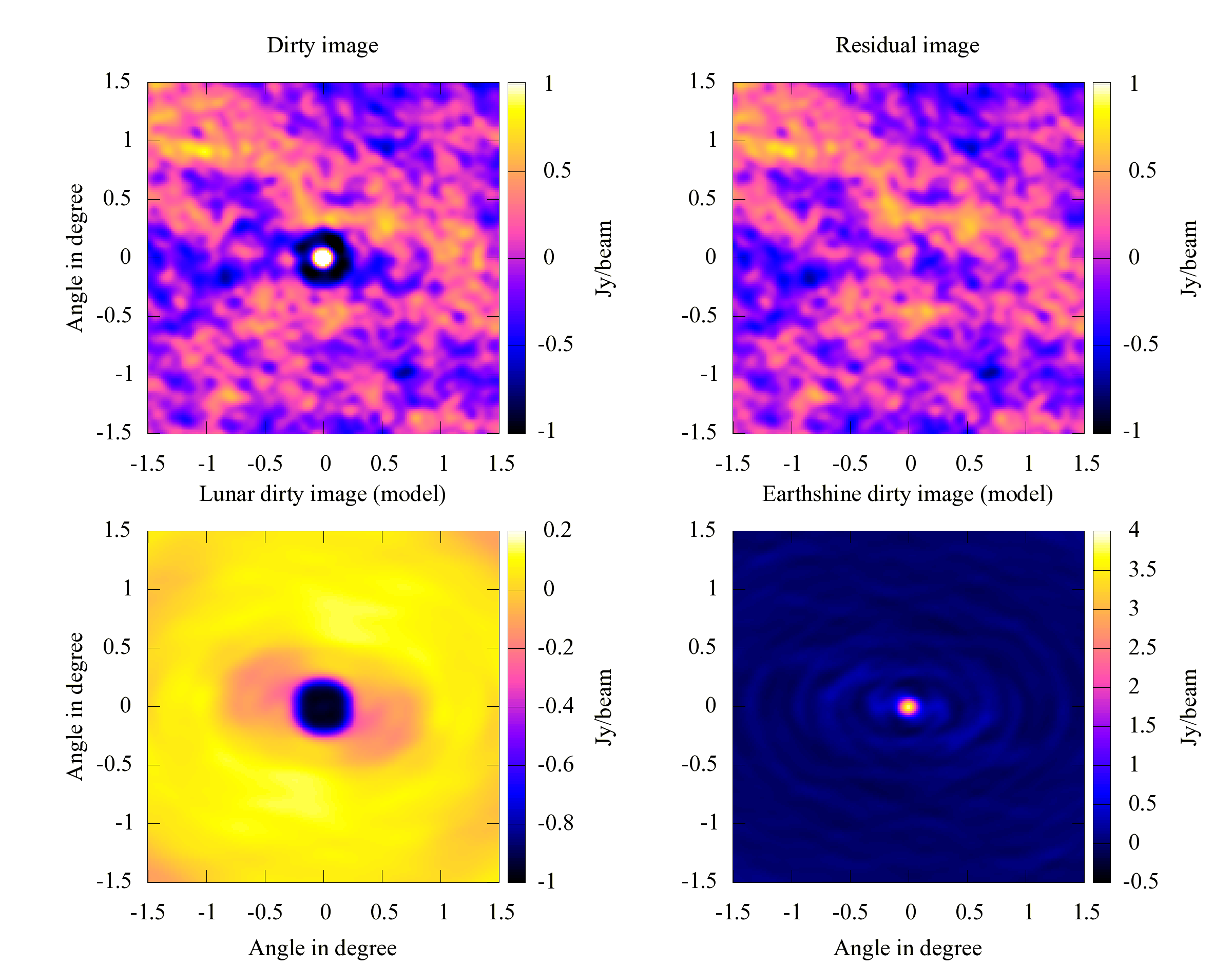}
\caption{Plot demonstrating the mitigation of reflected RFI (Earthshine) in lunar images. Top left panel shows a dirty image of the Moon at $68$~MHz contaminated by Earthshine. Bottom-left panel shows the reconstructed dirty image of the Moon with Earthshine removed. Bottom-right panel shows the reconstructed dirty image of Earthshine only, and top-right panel shows the residuals of model fitting.\label{fig:earthshine_mitigation}} 
\end{figure*}

Figure \ref{fig:earthshine_mitigation} demonstrates our modeling procedure on a dirty image of the Moon made with a subband strongly contaminated by reflected Earthshine. The top left panel shows the dirty image of the Moon $\mymat{D}$. The bright positive emission at the center of the lunar disc is due to reflected Earthshine. The bottom left and right panels show the reconstruction of the dirty images of the lunar disc alone (given by $\widehat{S}_{m}(\mymat{M}\ast\mymat{P}$), and that of reflected Earthshine alone (given by $\widehat{S}_{es}\mymat{P}$). The top right panel shows an image of the residuals of fit $\mymat{R}$. The residual images show non-thermal systematics with spatial structure. This is expected as the noise $\mymat{N}$ is dominated by sidelobe confusion and residuals of 3C144. Nevertheless, to first order, we have isolated the effect of reflected Earthshine from lunar images, and can now estimate the background temperature spectrum independently.

\subsection{Background temperature estimation}
\label{subsec:final_estimation}
\begin{figure*}
\centering
\includegraphics[width=\linewidth]{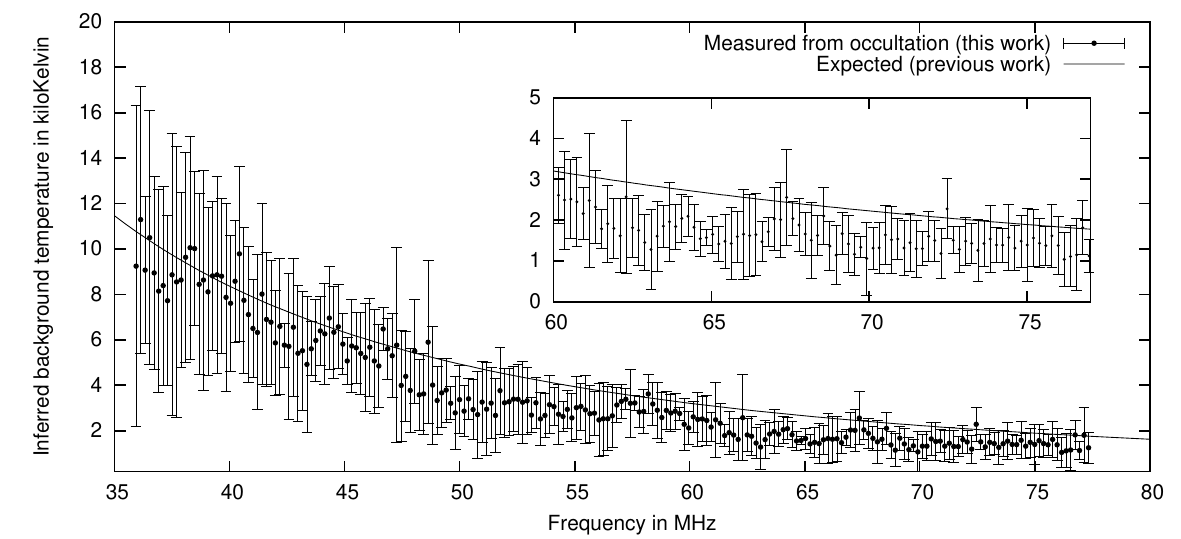}
\caption{Plot showing the inferred background temperature occulted by the Moon. At each channel, we estimate $7$ background temperature values using $7$ one-hour synthesis images of the Moon. Plotted are the mean and standard deviation of the $7$ temperatures measured in each channel. \label{fig:tback}}
\end{figure*}

As shown in Section \ref{subsec:resp_to_occultation}, the estimated flux of the lunar disc $\widehat{S}_\textrm{m}$ is a measure of the brightness temperature contrast between the Moon and the background: $T_M-T_B$. Using Equations \ref{eqn:tmoon} and \ref{eqn:stot}, we get the estimator for the background-temperature spectrum as

\begin{equation}
\label{eqn:tback}
\widehat{T_B}(\nu) = 230+160\left(\frac{\nu\,\,\textrm{MHz}}{60}\right)^{-2.24}-\frac{10^{-26}c^2\widehat{S}_{m}}{2k\Omega\nu^2}.
\end{equation}
Figure \ref{fig:tback} shows the estimates $\widehat{T_B}(\nu)$ computed from Equation \ref{eqn:tback} and lunar disc flux estimates $\widehat{S}_{m}(\nu)$ from the fitting procedure described in Section \ref{subsec:lls_fit}.\\
 
For the data in Fig. \ref{fig:tback}, we split our $7$~hour synthesis into $7$ one-hour syntheses. Hence, for each frequency channel ($195$~kHz wide), we obtain $7$ estimates of the occulted brightness temperature. In figure \ref{fig:tback} we plot the mean (black points) and standard deviation (error bars) of $7$ temperature estimates at each frequency channel. The actual uncertainties on the estimates (given by Equation \ref{eqn:cov}) is significantly smaller than the standard deviation, and hence, we do not show them on this plot. We have applied two minor corrections to the data in Fig. \ref{fig:tback}: (i) Since the Moon moves with respect to the primary-beam tracking point during the synthesis, we apply a net ($7$~hour averaged) primary-beam correction for each frequency using a simple analytical primary-beam model (array factor), and (ii) we bootstrapped the overall flux scale to the 3C123 scale from \citet{perley_scale}. These two corrections only lead to a marginal flattening of the estimated background spectrum, but we include them nevertheless for completeness.\\

Though the Moon moves by about $0.5$~deg per hour, the large hour-to-hour variation seen in our data (error bars in Fig. \ref{fig:tback}) is likely not intrinsic to the sky, and mostly emanates from sidelobe-confusion noise on short baselines where most of the lunar flux lies. Due to a lack of accurate models for bright in-field sources (3C144 for instance) and the complex resolved Galactic structure (close to the Galactic plane), we are unable to mitigate this confusion with current data. Fig. \ref{fig:tback} also shows the expected Galactic spectrum from available sky models from \citet{dacosta} (solid line) which may be approximated by a power law with spectral index $\alpha = -2.364$ and a temperature of $3206$~K at $60$~MHz\footnote{The \citet{dacosta} sky model is a composite of data from various surveys, and as such, may suffer uncertainties due to scaling and zero-point offset in these surveys.}. The inferred background spectrum from our data is best fitted by a power law of index $\alpha=-2.9$ with a temperature of $2340$~K at $60$~MHz. If we assume the background spectrum of \citet{dacosta} to be true, then our data imply a lunar brightness temperature (thermal emission) of around $1000$~K with a lunar albedo at $7$\%, or a lunar albedo of around $30$\% with a lunar brightness temperature to $230$~K. The nominal values for the lunar albedo ($7$\%) and thermal emission ($230$~K black body) have been taken from measurements at higher frequencies  ($\nu>200$~MHz). The lowest frequency measurements of the lunar thermal emission that we are aware of is the one at $178$~MHz by \citet{baldwin1961}, where the authors did not find significant deviation from the nominal value of $230$~K. The penetration depth into the lunar regolith for radiation with wavelength $\lambda$ is $\sim100\lambda$ \citep{baldwin1961}. Though significant uncertainties persist for penetration depth estimates, using the above value, the penetration depth in our observation bandwidth varies between $375$ and $860$~meter. Lunar regolith characteristics have not been constrained at these depths so far. Nevertheless, given the high amount ($10-20$\%) of systematics in our estimate of the background temperature spectrum (due to confusion from unmodeled flux in the field), any suggestions of evolution of lunar properties at lower frequencies (larger depths) at this point is highly speculative. In any case, we expect future observations proposed in Section \ref{subsec:way_forward} to resolve the current discrepancy we observe in the data.

\subsection{Earthshine estimation}
\begin{figure*}
\centering
\includegraphics[width=\linewidth]{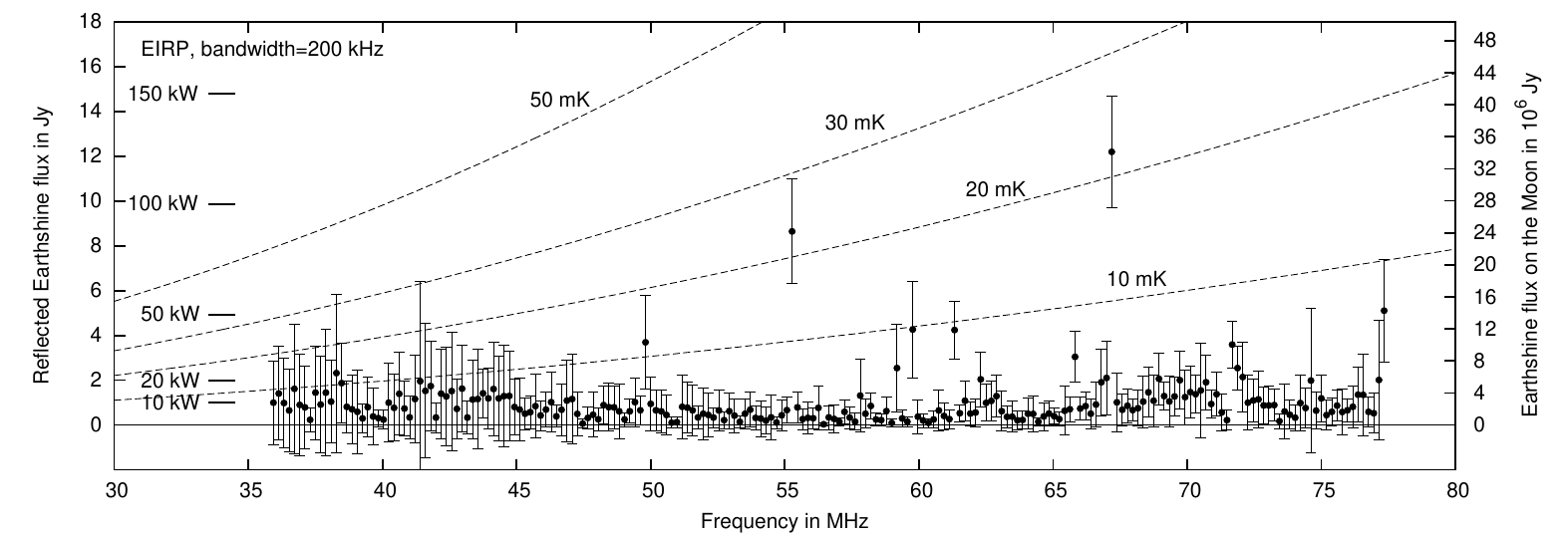}
\caption{Plot showing the reflected Earthshine (left-hand y axis) and Earthshine as seen by an observer on the Moon (right-hand y axis). At each frequency channel ($195$~kHz wide), we measure $7$ values of Earthshine each from a one hour synthesis images ($7$~hours in total). Plotted are the mean and standard deviation of the $7$ Earthshine values at each frequency channel. Also shown are $5$ EIRP levels of the Earth in a $195$~kHz bandwidth. The dotted lines denote the flux level in a single dipole for different brightness temperatures of $10,20,30,$~and $50$~mK. \label{fig:earthshine_estimate}}
\end{figure*}

Given the Earth-Moon distance $D$, and the effective back-scattering cross section of the Moon \footnote{Also called the Radar Cross Section (RCS) in Radar literature.} $\sigma_m$, we can convert the estimated values $\widehat{S}_{es}(\nu)$, to the incident Earthshine flux as seen by an observer on the Moon, $S_{inc}(\nu)$:
\begin{equation}
\label{eqn:friis}
S_{inc}(\nu) = \frac{\widehat{S}_{es}(\nu)\,\,4\pi D^2}{\sigma_{m}}.
\end{equation}
Following \citet{evans1969}, and since the Moon is large compared to a wavelength, its scattering cross section is independent of frequency, and equals its geometric cross section times the albedo:
\begin{equation}
\label{eqn:rcs}
\sigma_{m} = 0.07\,\pi R_{m}^2.
\end{equation}
Using Equation \ref{eqn:rcs} in Equation \ref{eqn:friis} gives
\begin{equation}
S_{inc}(\nu) = \frac{4}{0.07} \widehat{S}_{es}(\nu) \left(\frac{D}{R_{m}}\right)^2.
\end{equation}
Using, $D=384000$~km and $R_{m}=1738$~km, we get
\begin{equation}
S_{inc}(\nu) \approx 2.8\times 10^6\,\, \widehat{S}_{es}(\nu)\,\,\, \textrm{Jy}.
\end{equation}
Furthermore, $S_{inc}$ (in Jy) can be converted to the effective isotropic radiated power (EIRP) by a transmitter on the Earth within our channel width of $\Delta \nu\,\,\textrm{Hz}$ according to:
\begin{equation}
\textrm{EIRP}(\nu) = 4\pi \Delta \nu D^2S_{inc}(\nu)\,10^{-26} \,\,\, \textrm{Watt},
\end{equation}
Figure \ref{fig:earthshine_estimate} shows the best estimates $\widehat{S}_{es}(\nu)$ (left-hand y-axis), and the corresponding values of Earthshine flux as seen from the Moon $S_{inc}(\nu)$ (right-hand y-axis). The corresponding EIRP levels of transmitters on the Earth in a $200$~kHz bandwidth are also indicated. As in Fig. \ref{fig:tback}, we have plotted the mean and standard deviation of Earthshine estimates obtained from $7$ one-hour syntheses.\\

Estimates of $S_{inc}(\nu)$ in Fig. \ref{fig:earthshine_estimate} form critical inputs to proposed Moon-based dark ages and Cosmic Dawn experiments, such as DARE \citep{dare}. Due to the high-risk nature of space mission, we will conservatively assume that such experiments must attain systematic errors in their antenna temperature spectrum of about $1$~mK or lower. Our estimates from Fig. \ref{fig:earthshine_estimate} may then be converted to a minimum Earth-isolation that such experiments must design for. We present these estimates for median Earthshine values in different frequency bins in Table \ref{tab:dare}.
\begin{table}
\begin{tabular}{ccc}
\label{tab:dare}
{\bf Frequency (MHz)} & {\bf Mean }$\mathbf{S_{inc}}~{\bf MJy}$ & {\bf Isolation (dB)} \\ \hline \\
35-45 & 3.6 & 73  \\
45-55 & 2 & 78\\
55-65 & 2.7 & 68\\
65-75 & 4.3 & 69\\ \hline
\end{tabular}
\caption{Minimum Earth-isolation required for Moon-based dark ages and Cosmic Dawn experiments to achieve an Earthshine temperature lower than $1$~mK}
\end{table}
The values in Fig. \ref{fig:earthshine_estimate} may also be re-normalized to low frequency radio astronomy missions to other solar system locations. For instance, the Earth-Sun Lagrange point L$_2$ is about $3.9$ times further than the Moon, and hence the corresponding values for $S_{inc}$ are about $15$~times lower, giving a minimum Earth-isolation that is lower that the values in Table \ref{tab:dare} by about $12$~dB.\\

Fig. \ref{fig:earthshine_estimate} also shows the Moon-reflected Earthshine flux in a single dipole on the Earth corresponding to sky averaged brightness temperatures of $10,20,30,$ and $50$~mK (dashed lines). Since reflected Earthshine (from the Moon) is within $20$~mK of single-dipole brightness temperature, we expect the presence of the Moon in the sky to not be a limitation to current single-dipole experiments to detect the expected $100$~mK absorption feature \citep{pritchard2010} from Cosmic Dawn with $5\sigma$ significance.\\

Earthshine may also be reflected from man-made satellites in orbit around the Earth. Since single-dipole experiments essentially view the entire sky, the aggregate power scattered from all visible satellites may pose a limitation in such experiments. While a detailed estimation of such contamination is beyond the scope of this paper, we now provide approximate numbers. The strength of reflected earthshine scales as $\sigma d^{-4}$ where $d$ is the distance to the scattering object, and $\sigma$ is its back-scattering cross section. Due to the $d^{-4}$ scaling, we expect most of the back-scattered Earthshine to come from Low Earth Orbit (LEO) satellites which orbit the Earth at a height of $400-800$~km. If we assume (i) a back-scattering geometry, and (ii) that the satellite views the same portion of the Earth as the Moon, then we conclude that the back-scattering cross section of a satellite that scatters the same power into a single dipole as the Moon ($1-2$~Jy) to be $0.8$~m$^2$ at $400$~km height and $12.5$~m$^2$ at $800$~km height. We expect the majority of the scattered power to come from large satellites and spent rocket stages as their sizes are comparable to or larger than a wavelength. The myriad smaller ($<10$~cm) space debis, though numerous, are in the Rayleigh scattering limit and may be safely ignored\footnote{This is because in the Rayleigh scattering limit, the cross section of an object of dimension $x$ scales as $x^6$}. Moreover the differential delay in back-scattering from different satellites is expected to be sufficient to decorrelate RFI of even small bandwidths of $10$~kHz. Hence, we can add the cross sections of all the satellites to calculate an effective cross section. Assuming a $10$~Jy back-scatter power limit for reliable detection of the cosmic signal with a single dipole \footnote{This corresponding to $20$~mK of antenna (single dipole) temperature at $65$~MHz--- required for a $5\sigma$ detection of the Cosmic Dawn absorption feature}, we conclude that the effective cross section of satellites (at $800$~km), should be lower than $80$~m$^2$. For a height of $400$~km, we arrive at a total visible satellite cross-section bound of just $5$~m$^2$. Since there are thousands of cataloged satellites and spent rocket stages in Earth orbit\footnote{See Figure II from \citet{debris}}, reflection from man-made objects in Earth-orbit may pose a limitation to Earth-based Cosmic Dawn experiments.

\subsection{Next steps}
\label{subsec:way_forward}
Due to the pilot nature of this project, we chose to observe the Moon when it was in a field that presented a high brightness contrast (close to the Galactic plane) facilitating an easy detection, and when the Moon reached its highest elevation in the sky as viewed by LOFAR. The latter choice put the Moon close to the bright source 3C144, and may have contributed to a large systematic error in the background temperature spectrum measurements (see Figure \ref{fig:tback}). In future observations we plan to mitigate this systematic using two complementary approaches: (i) choice of a better field (in the Galactic halo) for easier bright source subtraction, and (ii) exploiting the $12$~deg per day motion of the Moon to cancel weaker sources through inter-night differencing \citep{shaver1999, mckinley2013}. In practice, residual confusion may persist due to differential ionospheric and primary-beam modulation between the two nights. Our next steps involve evaluating such residual effects.\\ 

We will now assume perfect cancellation of sidelobe noise and compute the thermal uncertainty one may expect in such an experiment. Since the SEFD is lower towards the Galactic halo, we use the value of $28$~kJy \citep{lofar} rather than the ones derived in Section \ref{subsec:sensitivity_analysis}. Conservatively accounting for a sensitivity loss factor of $1.36$ due to the time variable station projection in a $7$~hour synthesis, we expect an effective SEFD of about $38$~kJy. Taking into account the fact that the LOFAR baselines resolve the Moon to different extents, we compute a thermal uncertainty in Moon-background temperature contrast measurement (via inter-night differencing) in a $1$~MHz bandwidth of of $5.7$~K. The poorer sensitivity of the occultation based technique as compared to a single-dipole experiment with the same exposure time ensues from LOFAR's low snapshot filling factor for baselines that are sensitive to the occultation signal ($<100\lambda$). In contrast, the Square Kilometer Array Phase-1 (SKA1) will have a low frequency aperture array ($50-350$~MHz) with a filling factor of $90$ per-cent in its $450$-meter core\footnote{Based in the SKA1 system baseline design document: https://www.skatelescope.org/home/technicaldatainfo/key-documents/}. Lunar occultation observations with SKA1-Low can thus yield a significant detection of the $21$-cm signal from Cosmic Dawn in a reasonable exposure time (several hours).\\

In the near-term with LOFAR, a background spectrum with an uncertainty of few Kelvin, will place competitive constraints on Galactic synchrotron emission spectrum at very low frequencies ($<80$~MHz). Additionally, since the thermal emission from the Moon at wavelength $\lambda$ comes primarily from a depth of $\sim 100\lambda$ \citep{baldwin1961},  such accuracies may also provide unprecedented insight into lunar regolith characteristics (albedo and temperature evolution) up to a depth of $\sim 1$~km.

\section{Conclusions and future work}
\label{sec:concl}
In this paper, we have presented a theoretical framework for estimating the spectrum of the diffuse radio sky (or the global signal) interferometrically using lunar occultation. Using LOFAR data, we have also demonstrated this technique observationally for the first time. Further refinement of this novel technique may open a new and exciting observational channel for measuring the global redshifted $21$-cm signal from the Cosmic Dawn and the Epoch of Reionization. We find the following:
\begin{enumerate}

\item We observationally confirm predictions that the Moon appears as a source with negative flux density ($-25$~Jy at $60$~MHz) in interferometric images between $35$ and $80$~MHz (see Fig. \ref{fig:moon_frames}) since its apparent brightness is lower than that of the background sky that it occults. Consequently, we find the apparent brightness temperature of the Moon to be sufficiently (up to $10-20$\% systematic measurement error on its flux) described by (i) its intrinsic $230$~K black-body emission as seen at higher frequencies \citep{krotikov1963,heiles1963}, (ii) reflected Galactic emission (Moon position dependent) which is about $160$~K at $60$~MHz with a spectral index of $-2.24$, and (iii) reflected Earthshine comprising of the reflected radio-frequency interference from the Earth. Lack of reliable estimates of Earthshine was the limiting factor in prior work done by \citet{mckinley2013}. \\

\item Lunar images in some frequency channels have a compact positive flux source at the center of the (negative flux) lunar disc. We attribute this compact source to be due to reflected Earthshine (Radio Frequency Interference), and observationally confirm predictions that the Earthshine reflection off the lunar regolith is mostly specular in nature. We demonstrated how this Earthshine can be independently measured using resolution afforded by LOFAR's long ($>100\lambda$) baselines. Consequently, reflected Earthshine is currently not a limiting factor for our technique.\\

\item Our Earthshine measurements between $35$ and $80$~MHz imply an Earth flux as seen from the Moon of $2$-$4\times 10^6$~Jy (frequency dependent), although this value may be as high as $35\times 10^6$~Jy in some isolated frequency channels. These values require Dark Ages and Cosmic Dawn experiments from a lunar platform to design for a nominal Earthshine isolation of better than $80$~dB to achieve their science goals (assuming a conservative upper bound for RFI temperature in a single dipole of $1$~mK).\\

\item For Earth-based Cosmic Dawn experiments, reflected RFI from the Moon results in an antenna temperature (single dipole) of less than $20$~mK in the frequency range $35$ to $80$~MHz, reaching up to $30$~mK in isolated frequency channels. This does not pose a limitation for a significant detection of the $100$~mK absorption feature expected at $65$~MHz ($z=20$). However, if the total cross-section of visible large man-made objects (satellite and rocket stages) exceeds $80$~m$^2$ at $800$~km height, or just $5$~m$^2$ at $400$~km height, then their aggregate reflected RFI will result in a single-dipole temperature in excess of $20$~mK--- a potential limitation for Earth-based single-dipole cosmic dawn experiments.\\

\item We plan to mitigate the current systematic limitations in our technique through (a) lunar observations in a suitable field away from complex and bright Galactic plane, and (b) inter-day differencing of visibilities to cancel confusion from the field while retaining the lunar flux (the Moon moves by about $12$~deg per day). We expect to reach an uncertainty of $\sim 10$~K in our reconstruction of the radio background. If successful, such a measurement may not only constrain Galactic synchrotron models, but also place unprecedented constraints on lunar regolith characteristics up to a depth of $\sim 1$~km. 

\end{enumerate}

\section*{Acknowledgments}
LOFAR, the Low Frequency Array designed and constructed by ASTRON, has facilities in several countries, that are owned by various parties (each with their own funding sources), and that are collectively operated by the International LOFAR Telescope (ILT) foundation under a joint scientific policy. HKV and LVEK acknowledge the financial support from the European Research Council under ERC-Starting Grant FIRSTLIGHT - 258942. We thank the computer group at the Kapteyn Institute for providing the Python modules that we used to render Figure \ref{fig:moon_field}. Chiara Ferrari acknowledges financial support by the {\it “Agence Nationale de la Recherche”} through grant ANR-09-JCJC-0001-01. 

\bibliographystyle{mn2e}
\bibliography{mybib.bib}

\appendix
\section{Interferometric response to a global signal}
\label{app:a}
This section provides the intermediate steps in the derivation of Equation \ref{eqn:global} from Equation \ref{eqn:me}. Equation \ref{eqn:me} is
\begin{equation}
\label{eqn:app_me}
V(\bar{u},\nu) = \frac{1}{4\pi}\int\,d\Omega\,\,T_{sky}(\bar{r},\nu)e^{-2\pi \mathrm{i}\bar{u}.\bar{r}}
\end{equation}
The exponential in the integrand can be cast in a spherical harmonic expansion as \footnote{Also called plane wave expansion, or Rayleigh's expansion after lord Rayleigh.} \citep{harrington}
\begin{equation}
\label{eqn:app_expansion}
e^{-2\pi\mathrm{i}\bar{u}.\bar{r}} = 4\pi\sum_{l=0}^{\infty}\sum_{m=-l}^{l}\mathrm{i}^l\mathcal{J}_l(2\pi|\bar{u}||\bar{r}|)Y_{lm}(\theta_r,\phi_r)Y^{\star}_{lm}(\theta_u,\phi_u)
\end{equation}
where $\mathcal{J}_l$ is the spherical Bessel function of the first kind of order $l$, $Y_{lm}$ are the spherical harmonics for mode $(l,m)$, $(|\bar{r}|,\theta_r,\phi_r)$ are the spherical co-ordinates of the direction vector $\bar{r}$, and $(|\bar{u}|,\theta_u,\phi_u)$ are the spherical co-ordinates of the baseline vector $\bar{u}$. \\

Using, $|\bar{r}|=1$, substituting Equation \ref{eqn:app_expansion} in Equation \ref{eqn:app_me}, and interchanging the order of integration and summation, we get
\begin{eqnarray}
V(\bar{u},\nu) &=& \sum_{l=0}^{\infty}\sum_{m=-l}^{l}\mathrm{i}^l\mathcal{J}_l(2\pi|\bar{u}|)Y_{lm}(\theta_u,\phi_u) \nonumber \\
&&\int d\Omega T_{sky}(\bar{r},\nu)Y^{\star}_{lm}(\theta_r,\phi_r)
\end{eqnarray}
The above integral is simply the spherical harmonic expansion of the sky brightness distribution: $T_{sky}^{lm}(\nu)$. This gives us
\begin{equation}
\label{eqn:app_finalme}
V(\bar{u},\nu) = \sum_{l=0}^{\infty}\sum_{m=-l}^{l}\mathrm{i}^l\mathcal{J}_l(2\pi|\bar{u}|)Y_{lm}(\theta_u,\phi_u)\,T_{sky}^{lm}(\nu) 
\end{equation}
Equation \ref{eqn:app_finalme} shows that the measured visibility on a given baseline is simply a weighted sum of the spherical harmonic coefficients of the sky brightness temperature distribution. The weights are a product of the baseline length dependent factor $\mathcal{J}_l(2\pi|\bar{u}|)$ and a baseline orientation dependent factor $Y_{lm}(\theta_u,\phi_u)$.\\

If the sky is uniformly bright (global signal only), then 
\begin{equation}
T_{sky}^{lm}(\nu) = \left\{ \begin{array}{cl} T_B(\nu) & l=m=0\\ 
					  0 & \mathrm{otherwise}
\end{array}\right\}
\end{equation}
Substituting this in Equation \ref{eqn:app_finalme}, we get
\begin{equation}
V(\bar{u}) =  T_B\,\mathcal{J}_0(2\pi|\bar{u}|) = T_B\,\frac{\sin(2\pi|\bar{u}|)}{2\pi|\bar{u}|}
\end{equation}
which is Equation \ref{eqn:global}.

\section{Reflected emission}
\label{app:b}
We compute the intensity of Galactic and Extragalactic emission reflected from the Moon using ray tracing with an assumption of specular reflection. Under this assumption, given the Earth-Moon geometry, every pixel on the lunar surface corresponds to a unique direction in the sky that is imaged onto the lunar pixel as seen by the telescope. The algorithm used in computation of reflected emission at each epoch is summarized below.
\begin{enumerate}
\item Define a HealPix \citep{healpix} grid (N=32) on the lunar surface. Given RA,DEC of the Moon and UTC extract all the pixels `visible' from the telescope location. Compute the position vector $\widehat{i}$, and surface normal vector $\widehat{n}$ for each pixel.\\

\item For each lunar pixel $\widehat{i}$, define the plane of incidence and reflection using two vectors: (a) normal $\widehat{n}$, and (b) vector $\widehat{t}=(\widehat{n}\times\widehat{r_p})\times\widehat{n}$ tangential to the surface.\\

\item For each lunar pixel $\widehat{i}$, the corresponding direction vector which images onto that pixel is then given by $\widehat{r} = (\widehat{i}.\widehat{t})\widehat{t} + (-\widehat{i}.\widehat{n})\widehat{n}$. \\

\item Re-grid the sky model from \citet{dacosta} on the grid points specified by vectors $\widehat{r}$. We use gridding by convolution with a Gaussian kernel since a moderate loss of resolution is not detrimental to our computations. The value of each pixel in the regridded map $t^{\widehat{i}}_I(\nu)$ gives the temperature of incident radiation from direction $\widehat{r}$ on the corresponding pixel $\widehat{i}$ on the Moon at frequency $\nu$. The subscript $I$ denotes that this is the incident intensity.\\

\item The temperature of reflected emission from each pixel $\widehat{i}$ is then given by 
\begin{equation}
t^{\widehat{i}}_R = \underbrace{0.07}_{\textrm{albedo}}\,\,\,\underbrace{\widehat{n}.\widehat{i}}_{\textrm{projection}}\,\,\,t^{\widehat{i}}_I
\end{equation}
where subscript $R$ denotes that this is the reflected intensity.

\item Cast the Moon pixel co-ordinates in an appropriate map-projection grid. We use the orthographic projection (projection of a sphere on a tangent plane). 

\end{enumerate}

\begin{figure}
\centering
\includegraphics[width=\linewidth]{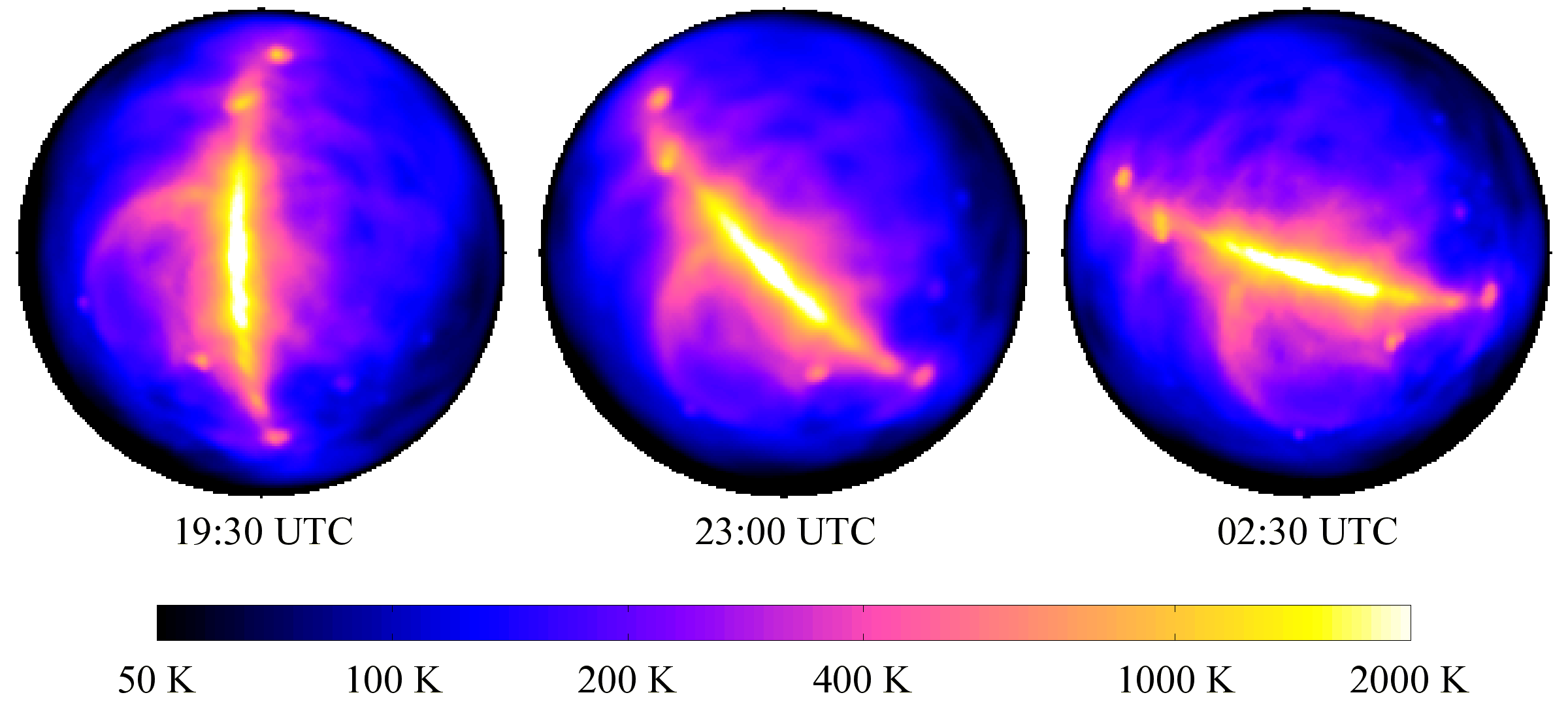}
\caption{Images showing the apparent brightness temperature of the lunar surface at $\nu=60$~MHz due to reflected Galactic and Extragalactic emission assuming specular reflection with a polarization independent albedo of $7$\%. The three panels correspond to $3$ epochs at the beginning, middle, and end of the synthesis, the data for which is presented in this paper \label{fig:app_3frames}}
\end{figure}

Figure \ref{fig:app_3frames} shows images of the computed apparent temperature of the lunar surface at $\nu=60$~MHz due to reflection of Galactic and Extragalactic emission. We only show images for $3$ epochs: beginning, middle, and end of our synthesis for which we presented data in this paper. The disc-averaged temperature in the images is $\approx160$~K (at $60$~MHz). Most of the apparent temporal variability (rotation) in the images is primarily due to the change in parallactic angle. The time variability of the disc-averaged temperature is $\sim 1$~K, and is discounted in subsequent analysis.

\section{Angular size of reflected Earthshine}
\label{app:c}
\begin{figure}
\centering
\includegraphics[width=\linewidth]{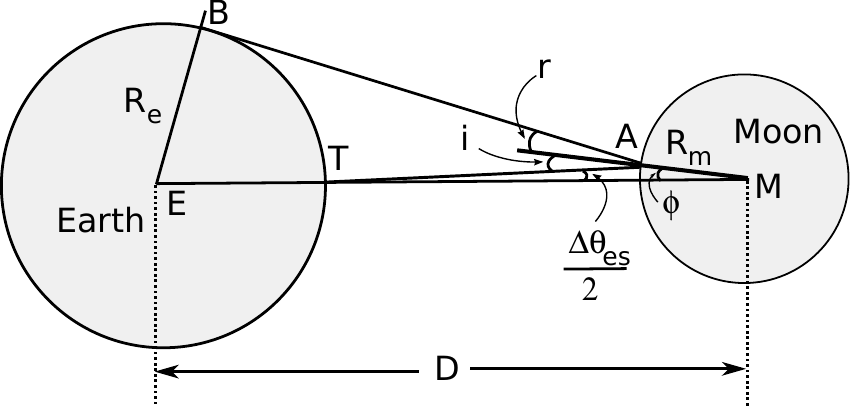}
\caption{Not-to-scale schematic of the Earth-Moon geometry used in the calculation of the angular size of reflected Earthshine\label{fig:emg}}
\end{figure}
As shown in Figure \ref{fig:emg}, we compute the angular size of reflected Earthshine $\Delta\theta_{es}$ by tracing the critical ray that emanates from the tangent point on the Earth (B), undergoes specular reflection on the lunar surface at A, and enters the telescope at T. The angle of incidence and reflection are then given by
\begin{equation}
\label{eqn:c1}
i = r = \phi + \frac{\Delta \theta_{es}}{2}
\end{equation}
Using $\sin(x) \approx x$ for $x\ll 1$ and applying the sine rule in triangle AMT, we get
\begin{equation}
\label{eqn:c2}
\frac{\phi}{D-R_m-R_e} \approx  \frac{\Delta\theta_{es}/2}{R_m},
\end{equation}
where we have approximated the length of segment AT by $D-R_m-R_e$.
Similarly, sine rule in triangle ABE gives
\begin{equation}
\label{eqn:c3}
\frac{i+r}{R_e} = \frac{2i}{R_e} = \frac{\sin(\frac{\pi}{2})}{D-R_m}
\end{equation}
Eliminating $i$ and $\phi$ between Equations \ref{eqn:c1}, \ref{eqn:c2}, and \ref{eqn:c3} we get
\begin{equation}
\Delta\theta_{es} = \frac{R_mR_e}{(D-R_m)(D-R_e)},
\end{equation}
which under the assumptions $D\gg R_m$ and $D\gg R_e$ yields
\begin{equation}
\Delta\theta_{es} \approx \frac{R_mR_e}{D^2},
\end{equation}
which is Equation \ref{eqn:spot}

\end{document}